\documentclass[aps,prc,amsfonts,preprintnumbers,superscriptaddress,showpacs,nofootinbib]{revtex4}
\pdfoutput=1
\usepackage{graphics}
\usepackage{amsmath}
\usepackage{amssymb}
\usepackage{psfrag}
\usepackage{epsfig}
\usepackage{epsf}
\usepackage{float}
\usepackage{slashed}
\allowdisplaybreaks

\newcommand{\fet}[1]{\mbox{\boldmath $#1$}}

\newcommand{\beq}{\begin{equation}}
\newcommand{\eeq}{\end{equation}}
\newcommand{\beqa}{\begin{eqnarray}}
\newcommand{\eeqa}{\end{eqnarray}}
\newcommand{\nn}{\nonumber \\ }

\begin{document}

\title{Three-nucleon force at large distances: Insights from chiral
  effective field theory and the large-$N_c$ expansion}

\author{E.~Epelbaum}
\email[]{Email: evgeny.epelbaum@rub.de}
\affiliation{Institut f\"ur Theoretische Physik II, Ruhr-Universit\"at Bochum,
  D-44780 Bochum, Germany}
\author{A.~M.~Gasparyan}
\email[]{Email: ashot.gasparyan@rub.de}
\affiliation{Institut f\"ur Theoretische Physik II, Ruhr-Universit\"at Bochum,
  D-44780 Bochum, Germany}
\affiliation{SSC RF ITEP, Bolshaya Cheremushkinskaya 25, 117218
  Moscow, Russia}
\author{H.~Krebs}
\email[]{Email: hermann.krebs@rub.de}
\affiliation{Institut f\"ur Theoretische Physik II, Ruhr-Universit\"at Bochum,
  D-44780 Bochum, Germany}
\author{C.~Schat}
\email[]{Email: carlos.schat@gmail.com}
\affiliation{Departamento de F\'{i}sica, FCEyN,
Universidad de Buenos Aires and IFIBA, CONICET,  Ciudad Universitaria,
Pab. 1, (1428) Buenos Aires, Argentina}
\date{\today}

\begin{abstract}
We confirm the claim of Ref.~\cite{Phillips:2013rsa} that $20$ operators are sufficient to represent the most
general local isospin-invariant three-nucleon force and derive explicit relations
between the two sets of operators suggested in Refs.~\cite{Phillips:2013rsa} and \cite{Krebs:2013kha}. 
We use the set of $20$ operators to discuss the chiral expansion of the long- and intermediate-range parts of
the three-nucleon force up to next-to-next-to-next-to-next-to-leading
order in the standard formulation without explicit $\Delta$(1232)
degrees of freedom.  We also address implications of the large-$N_c$ expansion in QCD for
the size of the various three-nucleon force contributions. 
\end{abstract}


\maketitle

\vspace{-0.2cm}

\section{Introduction}
\def\theequation{\arabic{section}.\arabic{equation}}
\label{sec:intro}

The three-nucleon force (3NF) has been a subject of intense research in nuclear
physics for many decades, see Refs.~\cite{KalantarNayestanaki:2011wz,Hammer:2012id} for recent
review articles. Explicit calculations have demonstrated that 3NFs
have significant effects in spectra and other
properties of light and medium-mass nuclei, see 
Refs.~\cite{Navratil:2007we,Pieper:2007ax,Hebeler:2009iv,Epelbaum:2009zsa,Maris:2011as,Roth:2011ar} for a selection of recent
studies along these lines. Three-body continuum provides an even more
clean and detailed testing ground for 3NFs. In particular, one expects that 3NF will
resolve several puzzles observed in nucleon-deuteron (Nd)
scattering at low energy such as the underprediction of the vector
analyzing power in elastic Nd scattering known as the  $A_y$ puzzle and the
discrepancy observed for the cross section in the so-called symmetric space star
configuration of the deuteron break up, see \cite{KalantarNayestanaki:2011wz} and references
therein. Moreover, effects of the 3NF in Nd scattering are
expected to become more prominent at energies above $E_{\rm lab} \sim
100$ MeV, where large deviations between calculations based on modern
high-precision potential models and 
experimental data are observed \cite{Gloeckle:1995jg}. 
The currently available phenomenological
3NF models are unable to explain these differences in
elastic scattering and deuteron breakup reactions which especially
applies to spin-dependent observables
\cite{KalantarNayestanaki:2011wz}. The much worse understanding of the
spin structure of the 3NF compared to the two-nucleon force is, to a
large extent, due to  a much richer operator structure of the 3NF,
a large computational effort needed to solve the three-body Faddeev
equations and a considerably more scarce data base in the three-nucleon
sector. Further progress in this field clearly requires guidance from
the theory in form of lattice QCD \cite{Beane:2012vq},
chiral effective field theory (EFT)  \cite{Epelbaum:2008ga} 
or large-$N_c$ expansion in QCD \cite{Phillips:2013rsa}. 

In the present work, we mainly focus on  the description of
the 3NF within the chiral expansion. Chiral EFT provides a systematic
and model independent approach to nuclear forces which relies on the
symmetries of QCD such as especially the spontaneously broken
approximate chiral symmetry, see Ref.~\cite{Epelbaum:2012vx} for an
introduction and  Refs.~\cite{Epelbaum:2008ga,Machleidt:2011zz} for
recent review articles on this subject. The first nonvanishing
contributions to the 3NF appear at next-to-next-to-leading order (N$^2$LO) in
the chiral expansion \cite{Epelbaum:2002vt} \footnote{This statement applies to
  energy-independent formulations of nuclear potentials.} and are given by
tree-level diagrams representing two-pion ($2\pi$) exchange, one-pion
exchange-contact and purely short-range contact interactions. 
The resulting 3NF at N$^2$LO has been extensively 
explored in few- and many-body studies during the past decade. Leading
corrections to the 3NF emerge at
next-to-next-to-next-to-leading order (N$^3$LO) from one-loop diagrams constructed
from the lowest-order vertices in the effective Lagrangian and have
been worked out recently \cite{Ishikawa:2007zz,Bernard:2007sp,Bernard:2011zr}. 
The very first calculations of 
nucleon-deuteron scattering observables using the 3NF up to N$^3$LO indicate that
the N$^3$LO corrections are rather weak and will not provide solution
to the low-energy puzzles mentioned above \cite{Witala:2013kya,Golak:2014ksa}.  In
fact, given that the lowest-order pion-nucleon vertices in the effective chiral
Lagrangian do \emph{not} receive contributions from the $\Delta$(1232)
resonance, one might expect large corrections from subleading,
i.e. next-to-next-to-next-to-next-to-leading order (N$^4$LO)
terms. The corresponding long- and intermediate-range contributions
are driven by the low-energy constants (LECs) $c_i$ which accompany
subleading pion-nucleon vertices. The LECs $c_{2,3,4}$ are, to a large
extent, governed by the $\Delta$ isobar and known to be numerically
rather large. This observation provides a strong motivation to extend
the derivation of the 3NF to N$^4$LO in the chiral expansion. In
Refs.~\cite{Krebs:2012yv,Krebs:2013kha}, this task was accomplished for the longest-range
$2\pi$-exchange and the intermediate-range two-pion-one-pion ($2\pi$-$1\pi$)
exchange and ring topologies, respectively. In order to be able to
address the convergence of the chiral expansion in a meaningful way, a
set of $22$ operators parametrizing the most general operator structure of a
local 3NF was suggested in Ref.~\cite{Krebs:2013kha}. By applying all
possible permutations of the nucleon labels, these operators give rise
to $89$ structures in the 3NF.  The structure of the 3NF was also
analyzed independently in Ref.~\cite{Phillips:2013rsa}
in the context of  the large-$N_c$ expansion in QCD. It was found in
this work that only $80$ independent structures appear in a most general
parametrization of a local 3NF. 

In this paper we confirm the conclusion of Ref.~\cite{Phillips:2013rsa} that the 
number of independent operators for the local three-nucleon force can be
reduced to $80$ and give a set of $20$ operators which generate
these $80$ structures upon performing all possible permutations.   
Since these findings affect the results for the structure functions
in coordinate space plotted in Figs.~4-8 of  Ref.~\cite{Krebs:2013kha}, we 
re-analyze the chiral expansion of the long- and intermediate-range
topologies employing the new set of $20$ operators.  We also correct
for a numerical error we found  in the Fourier transformation of the 
$2\pi$-exchange in Ref.~\cite{Krebs:2013kha} which resulted in enhanced size of certain structure
functions. Notice that only figures but none of the expressions given
in that work are affected by the above-mentioned error. 
Finally, we discuss implications of the large-$N_c$ expansion in QCD for
the size of the various three-nucleon force contributions. 

Our paper is organized as follows. In section \ref{sec:lagr} we
provide explicit relations between the redundant operators given
in  Ref.~\cite{Krebs:2013kha} and define a set of $20$ independent
operators both in coordinate and momentum spaces. Next, in section
\ref{sec:3nfcoordspace}, we show the results for the corresponding
structure functions of the $2\pi$-, $2\pi$-$1\pi$-exchange and ring
topologies in the equilateral triangle configuration and discuss
convergence of the chiral expansion. Section \ref{sec:largeNC}
addresses 
implications of the large-$N_c$ expansion on the size of the various
terms. Finally, the main results of this study are summarized  in section \ref{sec:summary}.

\section{Local three-nucleon forces}
\def\theequation{\arabic{section}.\arabic{equation}}
\label{sec:lagr}

A general local three-nucleon force in momentum space can be written in a form
 $$
V_{\rm 3N} =  \sum_i {O}_i(\vec\sigma_1,\vec \sigma_2, \vec\sigma_3,\fet
 \tau_1,\fet \tau_2, \fet \tau_3,\vec q_1, \vec q_3) \, F_i(q_1, q_3, \vec q_1\cdot \vec q_3)\,,
 $$
 where $\vec \sigma_i$ ($\fet \tau_i$) denote spin (isospin) Pauli
 matrices for the nucleon $i$ and  $\vec q_{i} = \vec p_i \,
' - \vec p_i$,  with $\vec p_i \, '$ and $\vec p_i$ being the final
and initial momenta of the nucleon $i$. Further, 
${O}_i$ are spin-momentum-isospin operators and the scalar
 structure functions $F_i$ depend on $q_1\equiv | \vec q_1|$,
 $q_3\equiv | \vec q_3|$ and the scalar product
 $\vec{q}_1\cdot\vec{q}_3$
or, equivalently, on $q_1$, $q_2$ and $q_3$. Here and in the
following, we require that the 3NF $V_{\rm 3N}$ is given in a symmetrized form with
respect to interchanging the nucleon labels. Assuming parity and
time-reversal invariance as well as isospin symmetry, a set of $89$
operators $O_i$ was suggested in Ref.~\cite{Krebs:2013kha}.
Alternatively,  $V_{\rm 3N}$ can be generated by $22$ operators upon
applying all possible permutations of the nucleon labels    
\beq
V_{\rm 3N} = \sum_{i=1}^{22}{\cal G}_i(\vec\sigma_1,\vec \sigma_2,
\vec\sigma_3,\fet \tau_1,\fet \tau_2, \fet \tau_3,\vec q_1, \vec q_3){\cal F}_i(q_1, q_3, \vec q_1\cdot \vec q_3)+5\,{\rm permutations}
\label{strf22}\,,
\eeq
where ${\cal F}_i$ denote the structure functions in  this
representation.  We show in Table~\ref{generalstr_old} both sets 
of the operators given in Ref.~\cite{Krebs:2013kha}. 
\begin{table}[t]
\begin{tabular}{|c|c|c|c|c|c|c|}
\hline
Generators ${\cal G}$ of 89 independent operators & $S$ & $A$ & $G_{12}$ & $G_{22}$ & $G_{11}$ & $G_{21}$\\
\hline
${\cal G}_{1} = 1$ & $O_1$ & 0 & 0 & 0 & 0& 0\\
\hline
${\cal G}_{2} =\fet \tau_1\cdot\fet \tau_3$ & $O_2$ & 0 & $O_3$ & $O_4$& $0$ & $0$\\
\hline 
${\cal G}_{3} =\vec{\sigma}_1\cdot\vec{\sigma}_3$ & $O_5$ & 0 & $O_6$ &  $O_7$& $0$ & $0$\\
\hline 
${\cal G}_{4} =\fet \tau_1\cdot\fet \tau_3\vec{\sigma}_1\cdot\vec{\sigma}_3$ &  $O_8$ & 0 &  $O_9$ &  $O_{10}$& $0$ & $0$\\
\hline 
${\cal G}_{5} =\fet \tau_2\cdot\fet \tau_3\vec{\sigma}_1\cdot\vec{\sigma}_2$ &  $O_{11}$ &   $O_{12}$ &  $O_{13}$&  $O_{14}$&   $O_{15}$ &   $O_{16}$\\
\hline 
${\cal G}_{6} =\fet \tau_1\cdot(\fet \tau_2\times\fet
\tau_3)\vec{\sigma}_1\cdot(\vec{\sigma}_2\times\vec{\sigma}_3)$ &
$O_{17}$ &  0 & 0 & 0 &  0 &  0\\
\hline 
${\cal G}_{7} =\fet \tau_1\cdot(\fet \tau_2\times\fet
\tau_3)\vec{\sigma}_2\cdot(\vec{q}_1\times\vec{q}_3)$ &
$O_{18}$ &  0 &  ${O}_{19}$&  $O_{20}$&  $0$ &  $0$\\
\hline 
${\cal G}_{8} =\vec{q}_1\cdot\vec{\sigma}_1\vec{q}_1\cdot\vec{\sigma}_3$ &  $O_{21}$ &   $O_{22}$ &   $O_{23}$ &  $O_{24}$&   $O_{25}$ &   $O_{26}$\\
\hline 
${\cal G}_{9} =\vec{q}_1\cdot\vec{\sigma}_3\vec{q}_3\cdot\vec{\sigma}_1$ &  $O_{27}$ &  0 &   $O_{28}$ &  $O_{29}$&  $0$&  $0$\\
\hline 
${\cal G}_{10} =\vec{q}_1\cdot\vec{\sigma}_1\vec{q}_3\cdot\vec{\sigma}_3$ &  $O_{30}$ &  0 &   $O_{31}$ &  $O_{32}$&  $0$&  $0$\\
\hline 
${\cal G}_{11} =\fet \tau_2\cdot\fet \tau_3\vec{q}_1\cdot\vec{\sigma}_1\vec{q}_1\cdot\vec{\sigma}_2$ &  $O_{33}$ &    $O_{34}$  &   $O_{35}$ &  $O_{36}$&   $O_{37}$ &    $O_{38}$ \\
\hline 
${\cal G}_{12} =\fet \tau_2\cdot\fet \tau_3\vec{q}_1\cdot\vec{\sigma}_1\vec{q}_3\cdot\vec{\sigma}_2$ &  $O_{39}$ &    $O_{40}$  &   $O_{41}$ &  $O_{42}$&   $O_{43}$ &    $O_{44}$ \\
\hline 
${\cal G}_{13} =\fet \tau_2\cdot\fet \tau_3\vec{q}_3\cdot\vec{\sigma}_1\vec{q}_1\cdot\vec{\sigma}_2$ &  $O_{45}$ &   $O_{46}$  &   $O_{47}$ &  $O_{48}$&   $O_{49}$ &    $O_{50}$ \\
\hline 
${\cal G}_{14} =\fet \tau_2\cdot\fet \tau_3\vec{q}_3\cdot\vec{\sigma}_1\vec{q}_3\cdot\vec{\sigma}_2$ &  $O_{51}$ &    $O_{52}$  &   $O_{53}$ &  $O_{54}$&   $O_{55}$ &    $O_{56}$ \\
\hline 
${\cal G}_{15} =\fet \tau_1\cdot\fet
\tau_3\vec{q}_2\cdot\vec{\sigma}_1\vec{q}_2\cdot\vec{\sigma}_3$ &
$O_{57}$ &   0  &   ${O}_{58}$ &  $O_{59}$&$0$&   $0$\\
\hline 
${\cal G}_{16} =\fet \tau_2\cdot\fet \tau_3\vec{q}_3\cdot\vec{\sigma}_2\vec{q}_3\cdot\vec{\sigma}_3$ &  $O_{60}$ &   $O_{61}$  &   $O_{62}$ &  $O_{63}$&  $O_{64}$ &    $O_{65}$\\
\hline 
${\cal G}_{17} =\fet \tau_1\cdot\fet \tau_3\vec{q}_1\cdot\vec{\sigma}_1\vec{q}_3\cdot\vec{\sigma}_3$ &  $O_{66}$ & $0$  &   $O_{67}$ &  $O_{68}$& $0$ &   $0$\\
\hline 
${\cal G}_{18} =\fet \tau_1\cdot(\fet \tau_2\times\fet \tau_3)\vec{\sigma}_1\cdot\vec{\sigma}_3\vec{\sigma}_2\cdot(\vec{q}_1\times\vec{q}_3)$ &  $O_{69}$ & 0  &   $O_{70}$ &  $O_{71}$& $0$ &   $0$\\
\hline 
${\cal G}_{19} =\fet \tau_1\cdot(\fet \tau_2\times\fet
\tau_3)\vec{\sigma}_3\cdot\vec{q}_1\vec{q}_1\cdot(\vec{\sigma}_1\times\vec{\sigma}_2)$
& $\; O_{72} \; $ &  $ \;  O_{73} \; $ &   $ \;  O_{74} \; $ &  $ \;  O_{75} \; $&  $ \;  O_{76} \; $ &   $ \;  O_{77} \;$\\
\hline 
${\cal G}_{20} =\fet \tau_1\cdot(\fet \tau_2\times\fet \tau_3)\vec{\sigma}_1\cdot\vec{q}_1\vec{\sigma}_2\cdot\vec{q}_1\vec{\sigma}_3\cdot(\vec{q}_1\times\vec{q}_3)$ &  $O_{78}$ &  $O_{79}$ &   $O_{80}$ &  $O_{81}$&  $O_{82}$ &   $O_{83}$\\
\hline 
${\cal G}_{21} =\fet \tau_1\cdot(\fet \tau_2\times\fet \tau_3)\vec{\sigma}_1\cdot\vec{q}_2\vec{\sigma}_3\cdot\vec{q}_2\vec{\sigma}_2\cdot(\vec{q}_1\times\vec{q}_3)$ &  $O_{84}$ & 0 &   $O_{85}$ &  $O_{86}$& $0$ &  $0$\\
\hline 
$\quad {\cal G}_{22} =\fet \tau_1\cdot(\fet \tau_2\times\fet
\tau_3)\vec{\sigma}_1\cdot\vec{q}_1\vec{\sigma}_3\cdot\vec{q}_3\vec{\sigma}_2\cdot(\vec{q}_1\times\vec{q}_3)
\quad$
& $O_{87}$ & $0$ &   ${O}_{88}$ &  ${O}_{89}$& $0$ &  $0$\\
\hline 
\end{tabular}
\caption{The set of 22 generating operators ${\cal G}_i$ and their
  relation to 89 operators ${O}_1,\dots,{O}_{89}$ suggested in
  Ref.~\cite{Krebs:2013kha}.  The operators ${O}_i$ are
  generated by application of one of the 6 functions $S, A, G_{11},
  G_{12}, G_{21}, G_{22}$ defined in the text on the corresponding operator ${\cal
    G}_j$.
\label{generalstr_old}}
\end{table}
The functions $S, A, G_{11},   G_{12}, G_{21}, G_{22}$ appearing in
this table refer to the corresponding irreducible representations
of the group $S_3$ and are defined via:
\beq
S({O})=\frac{1}{6}\sum_{P\in S_3}P{O}, \quad
A({O})=\frac{1}{6}\sum_{P\in S_3}(-1)^{w(P)}P{O}, 
\quad 
G_{ij}({O})=\frac{1}{3}\sum_{P\in S_3}{\cal D}_{ij}(P) PO, \quad
\mbox{with } i,j=1,2\,,
\eeq
where $w(P) =\pm 1$ for even/odd permutations and the matrices ${\cal D}$ for the
two-dimensional representation can be chosen e.g.~in the form 
\beq
\begin{array}{lll}
{\cal D}(())=\left(
\begin{array}{cc}
1&0\\
0&1
\end{array}
\right),
&{\cal D}((12))=\frac{1}{2}\left(
\begin{array}{cc}
1&\sqrt{3}\\
\sqrt{3}&-1
\end{array}
\right),
&{\cal D}((13))=\left(
\begin{array}{cc}
-1& 0\\
0 & 1
\end{array}
\right),\\
{\cal D}((23))=-\frac{1}{2}\left(
\begin{array}{cc}
-1&\sqrt{3}\\
\sqrt{3}&1
\end{array}
\right),
&{\cal D}((123))=-\frac{1}{2}\left(
\begin{array}{cc}
1&\sqrt{3}\\
-\sqrt{3}&1
\end{array}
\right),
&
{\cal D}((132))=-\frac{1}{2}\left(
\begin{array}{cc}
1&-\sqrt{3}\\
\sqrt{3}&1
\end{array}
\right), 
\end{array}
\eeq
see Ref.~\cite{Krebs:2013kha} for more details. 

As pointed out in Ref.~\cite{Phillips:2013rsa}, the number of independent operators for the
local three-nucleon force can be reduced from $89$ to $80$. 
This can be most easily seen by forming irreducible tensor operators
separately from the Pauli matrices and momenta and contracting them
with each other. More
precisely, we found that the operators $O_{78 \ldots 86}$ are
redundant and can be expressed in terms of the remaining
operators as follows: 
\beqa
O_{78}&=& \frac{1}{12} 
\left(q_1^4-4 q_1^2 \left(q_2^2+q_3^2\right)+q_2^4-4 q_2^2 
q_3^2+q_3^4\right) O_{17}+\frac{\sqrt{3}}{8} \left(q_1^2-q_3^2\right) 
O_{70}-\frac{1}{8} \left(q_1^2-2 q_2^2+q_3^2\right) 
O_{71}\nn
&+&\frac{1}{2} \left(q_1^2+q_2^2+q_3^2\right) 
O_{72}+\frac{1}{4} \left(q_3^2-q_1^2\right) O_{76}+\frac{1}{4 
\sqrt{3}}\left(q_1^2-2 q_2^2+q_3^2\right) O_{77}-2 O_{87},\nn
O_{79}&=& -\frac{1}{8 
\sqrt{3}}\left(q_1^2-2 q_2^2+q_3^2\right) O_{70}+\frac{1}{8} \left(q_3^2-q_1^2\right) 
O_{71}-\frac{1}{6} \left(q_1^2+q_2^2+q_3^2\right) 
O_{73}+\frac{1}{12} \left(q_1^2-2 q_2^2+q_3^2\right) 
O_{76}\nn
&+&\frac{1}{4 \sqrt{3}}\left(q_1^2-q_3^2\right) O_{77},\nn
O_{80}&=& -\frac{1}{8 \sqrt{3}}\left(q_1^2-q_3^2\right) \left(q_1^2-5 q_2^2+q_3^2
\right) O_{17}-\frac{3}{8} q_2^2 
O_{70}+\frac{\sqrt{3}}{8} \left(q_3^2-q_1^2
\right) O_{71}+\frac{3}{4} q_2^2 O_{74}+\frac{\sqrt{3}}{4} \left(q_1^2-q_3^2\right) 
O_{75}\nn
&-&\frac{\sqrt{3}}{2} q_2^2 
O_{76}+\frac{1}{2} \left(q_3^2-q_1^2\right)
O_{77}-\frac{1}{2}O_{88},\nn
O_{81}&=& \frac{1}{24} \left(q_1^4+5 q_1^2 
\left(q_2^2-2 q_3^2\right)-2 q_2^4+5 q_2^2 q_3^2+q_3^4\right) O_{17}+
\frac{\sqrt{3}}{8} \left(q_3^2-q_1^2\right) O_{70}+\frac{1}{8} \left(-2 q_1^2+q_2^2-2 q_3^2\right) 
O_{71}\nn
&+&\frac{\sqrt{3}}{4} 
\left(q_1^2-q_3^2\right) O_{74}+\frac{1}{4} \left(2 q_1^2-q_2^2+2 q_3^2\right) 
O_{75}+\frac{1}{2} 
\left(q_3^2-q_1^2\right) O_{76}+\frac{1}{2 \sqrt{3}}\left(-2 q_1^2+q_2^2-2 q_3^2\right) O_{77}-
\frac{1}{2}O_{89},\nn
O_{82}&=& \frac{1}{8} \left(q_1^2-q_3^2\right) 
\left(q_1^2-q_2^2+q_3^2\right) O_{17}+\frac{1}{8 \sqrt{3}}\left(-2 q_1^2+q_2^2-2 
q_3^2\right) O_{70}+\frac{1}{8} \left(q_1^2-q_3^2\right) 
O_{71}+\left(q_3^2-q_1^2\right) O_{72}\nn
&-&\frac{1}{3} \left(q_1^2-2 
q_2^2+q_3^2\right) O_{73}-\frac{ \sqrt{3}}{4} q_2^2 O_{74}+\frac{1}{4} \left(q_3^2-q_1^2
\right) O_{75}+\frac{1}{6} \left(q_1^2+4 q_2^2+q_3^2
\right) O_{76}+\frac{1}{2 
\sqrt{3}}\left(q_1^2-q_3^2\right) O_{77}
+\frac{ \sqrt{3} }{2}
O_{88},\nn
O_{83}&=&-\frac{1}{8 \sqrt{3}}\left(q_1^4+q_1^2 \left(q_2^2-2 q_3^2\right)-2 
q_2^4+q_2^2 q_3^2+q_3^4\right) O_{17}+\frac{1}{8} \left(q_1^2-q_3^2\right) 
O_{70}-\frac{\sqrt{3}}{8} 
q_2^2 O_{71}\nn
&+& \frac{1}{\sqrt{3}}\left(q_1^2-2 q_2^2+q_3^2\right) 
O_{72}+\frac{1}{\sqrt{3}}\left(q_3^2-q_1^2\right) O_{73}+\frac{1}{4} 
\left(q_3^2-q_1^2\right) O_{74}+\frac{1}{4 
\sqrt{3}}\left(-2 q_1^2+q_2^2-2 q_3^2\right) O_{75}\nn
&+&\frac{1}{2 \sqrt{3}}\left(q_1^2-q_3^2\right) 
O_{76}+\frac{1}{2} \left(q_1^2+q_3^2\right) 
O_{77}+\frac{\sqrt{3}}{2} O_{89},\nn
O_{84}&=& \frac{1}{12} \left(q_1^4-6 q_1^2 \left(q_2^2+q_3^2
\right)+q_2^4-6 q_2^2 q_3^2+q_3^4\right) O_{17}+\frac{\sqrt{3} }{4} 
\left(q_1^2-q_3^2\right) O_{70}-\frac{1}{4} \left(q_1^2-2 q_2^2+q_3^2\right) 
O_{71}\nn
&+&\left(q_1^2+q_2^2+q_3^2\right) O_{72}+\frac{1}{2} \left(q_3^2-q_1^2\right) 
O_{76}+\frac{1}{2 \sqrt{3}}\left(q_1^2-2 
q_2^2+q_3^2\right) O_{77}-2 O_{87},\nn
O_{85}&=& \frac{1}{4 \sqrt{3}}\left(q_1^2-q_3^2\right) \left(q_1^2-3 q_2^2+q_3^2
\right) O_{17}+\frac{1}{4} \left(q_1^2+q_2^2+q_3^2\right) 
O_{70}+\frac{1}{2} \left(q_1^2-2 q_2^2+q_3^2\right) 
O_{74}+\frac{\sqrt{3} }{2} \left(q_3^2-q_1^2\right) 
O_{75}\nn
&+&\frac{\sqrt{3} }{2} 
q_2^2 O_{76}+\frac{1}{2} \left(q_1^2-q_3^2\right)
O_{77}+O_{88},\nn
O_{86}&=& \frac{1}{12} 
\left(-q_1^4-3 q_1^2 \left(q_2^2-2 q_3^2\right)+2 q_2^4-3 q_2^2 
q_3^2-q_3^4\right) O_{17}+\frac{1}{4} \left(q_1^2+q_2^2+q_3^2\right) 
O_{71}+\frac{1}{2} \sqrt{3} \left(q_3^2-q_1^2\right) 
O_{74}\nn
&-&\frac{1}{2}\left(q_1^2-2 q_2^2+q_3^2\right) 
O_{75}+\frac{1}{2} \left(q_1^2-q_3^2\right) O_{76}+\frac{1}{2 
\sqrt{3}}\left(2 q_1^2-q_2^2+2 q_3^2\right) O_{77}+O_{89}.
\eeqa
One immediately observes from  Table~\ref{generalstr_old}  that the
operators ${\cal G}_{20}$ and ${\cal G}_{21}$ are
redundant. 

Here and in what follows, we adopt the new basis with 80 operators which can be generated by
$20$ operators given in momentum and coordinate spaces in Table~\ref{generalstr}.
\begin{table}[t]
\begin{tabular}{|c|c|}
\hline
Generators ${\cal G}$ in momentum space& Generators $\tilde{\cal G}$ in coordinate space\\
\hline
${\cal G}_{1} = 1$ & $\tilde  {\cal G}_{1} = 1$\\
${\cal G}_{2} =\fet \tau_1\cdot\fet \tau_3$ & $\tilde  {\cal G}_{2}= \fet \tau_1\cdot\fet \tau_3$ \\
${\cal G}_{3} =\vec{\sigma}_1\cdot\vec{\sigma}_3$ & $\tilde  {\cal G}_{3} =  \vec{\sigma}_1\cdot\vec{\sigma}_3$\\
${\cal G}_{4} =\fet \tau_1\cdot\fet \tau_3\vec{\sigma}_1\cdot\vec{\sigma}_3$ &  $\tilde  {\cal G}_{4} =  \fet \tau_1\cdot\fet \tau_3\, \vec{\sigma}_1\cdot\vec{\sigma}_3$\\
${\cal G}_{5} =\fet \tau_2\cdot\fet \tau_3\vec{\sigma}_1\cdot\vec{\sigma}_2$ & $\tilde  {\cal G}_{5} =  \fet \tau_2\cdot\fet \tau_3\, \vec{\sigma}_1\cdot\vec{\sigma}_2$\\
${\cal G}_{6} =\fet \tau_1\cdot(\fet \tau_2\times\fet
\tau_3)\vec{\sigma}_1\cdot(\vec{\sigma}_2\times\vec{\sigma}_3)$ &$\tilde  {\cal G}_{6} =   \fet \tau_1\cdot(\fet \tau_2\times\fet
\tau_3)\, \vec{\sigma}_1\cdot(\vec{\sigma}_2\times\vec{\sigma}_3)$\\
${\cal G}_{7} =\fet \tau_1\cdot(\fet \tau_2\times\fet
\tau_3)\vec{\sigma}_2\cdot(\vec{q}_1\times\vec{q}_3)$ &$\tilde  {\cal G}_{7} =   \fet \tau_1\cdot(\fet \tau_2\times\fet
\tau_3)\, \vec{\sigma}_2\cdot(\hat{r}_{12}\times\hat{r}_{23})$ \\
${\cal G}_{8}
=\vec{q}_1\cdot\vec{\sigma}_1\vec{q}_1\cdot\vec{\sigma}_3$ & $\tilde
{\cal G}_{8}  =   \hat{r}_{23}\cdot\vec{\sigma}_1\, \hat{r}_{23}\cdot\vec{\sigma}_3$\\
${\cal G}_{9}
=\vec{q}_1\cdot\vec{\sigma}_3\vec{q}_3\cdot\vec{\sigma}_1$ &  $\tilde
{\cal G}_{9}  =    \hat{r}_{23}\cdot\vec{\sigma}_3\, \hat{r}_{12}\cdot\vec{\sigma}_1$\\
${\cal G}_{10}
=\vec{q}_1\cdot\vec{\sigma}_1\vec{q}_3\cdot\vec{\sigma}_3$ &  $\tilde
{\cal G}_{10}  =  \hat{r}_{23}\cdot\vec{\sigma}_1\, \hat{r}_{12}\cdot\vec{\sigma}_3 $\\
${\cal G}_{11} =\fet \tau_2\cdot\fet
\tau_3\vec{q}_1\cdot\vec{\sigma}_1\vec{q}_1\cdot\vec{\sigma}_2$ &  
$\tilde  {\cal G}_{11}  =   \fet \tau_2\cdot\fet \tau_3\,
\hat{r}_{23}\cdot\vec{\sigma}_1\, \hat{r}_{23}\cdot\vec{\sigma}_2 $\\
${\cal G}_{12} =\fet \tau_2\cdot\fet
\tau_3\vec{q}_1\cdot\vec{\sigma}_1\vec{q}_3\cdot\vec{\sigma}_2$ &
$\tilde  {\cal G}_{12}  =  \fet \tau_2\cdot\fet \tau_3\,
\hat{r}_{23}\cdot\vec{\sigma}_1\, \hat{r}_{12}\cdot\vec{\sigma}_2 $\\
${\cal G}_{13} =\fet \tau_2\cdot\fet
\tau_3\vec{q}_3\cdot\vec{\sigma}_1\vec{q}_1\cdot\vec{\sigma}_2$ &
$\tilde  {\cal G}_{13}  =   \fet \tau_2\cdot\fet \tau_3\,
\hat{r}_{12}\cdot\vec{\sigma}_1\, \hat{r}_{23}\cdot\vec{\sigma}_2 $ \\
${\cal G}_{14} =\fet \tau_2\cdot\fet
\tau_3\vec{q}_3\cdot\vec{\sigma}_1\vec{q}_3\cdot\vec{\sigma}_2$ &
$\tilde  {\cal G}_{14}  =   \fet \tau_2\cdot\fet \tau_3\,
\hat{r}_{12}\cdot\vec{\sigma}_1\, \hat{r}_{12}\cdot\vec{\sigma}_2 $\\
${\cal G}_{15} =\fet \tau_1\cdot\fet
\tau_3\vec{q}_2\cdot\vec{\sigma}_1\vec{q}_2\cdot\vec{\sigma}_3$ &
$\tilde  {\cal G}_{15}  =    \fet \tau_1\cdot\fet
\tau_3\, \hat{r}_{13}\cdot\vec{\sigma}_1\, \hat{r}_{13}\cdot\vec{\sigma}_3 $\\
${\cal G}_{16} =\fet \tau_2\cdot\fet
\tau_3\vec{q}_3\cdot\vec{\sigma}_2\vec{q}_3\cdot\vec{\sigma}_3$ &
$\tilde  {\cal G}_{16}  =    \fet \tau_2\cdot\fet \tau_3\,
\hat{r}_{12}\cdot\vec{\sigma}_2\, \hat{r}_{12}\cdot\vec{\sigma}_3$\\
${\cal G}_{17} =\fet \tau_1\cdot\fet
\tau_3\vec{q}_1\cdot\vec{\sigma}_1\vec{q}_3\cdot\vec{\sigma}_3$ &
$\tilde  {\cal G}_{17}  =   \fet \tau_1\cdot\fet \tau_3\,
\hat{r}_{23}\cdot\vec{\sigma}_1\, \hat{r}_{12}\cdot\vec{\sigma}_3 $\\
${\cal G}_{18} =\fet \tau_1\cdot(\fet \tau_2\times\fet
\tau_3)\vec{\sigma}_1\cdot\vec{\sigma}_3\vec{\sigma}_2\cdot(\vec{q}_1\times\vec{q}_3)$
& $\tilde  {\cal G}_{18}  =   \fet \tau_1\cdot(\fet \tau_2\times\fet
\tau_3)\, \vec{\sigma}_1\cdot\vec{\sigma}_3\, \vec{\sigma}_2\cdot(\hat{r}_{12}\times\hat{r}_{23}) $\\
${\cal G}_{19} =\fet \tau_1\cdot(\fet \tau_2\times\fet
\tau_3)\vec{\sigma}_3\cdot\vec{q}_1\vec{q}_1\cdot(\vec{\sigma}_1\times\vec{\sigma}_2)$
& $\tilde
{\cal G}_{19}  =   \fet \tau_1\cdot(\fet \tau_2\times\fet
\tau_3)\, \vec{\sigma}_3\cdot\hat{r}_{23}\, \hat{r}_{23}\cdot(\vec{\sigma}_1\times\vec{\sigma}_2)$\\
$\quad {\cal G}_{20} =\fet \tau_1\cdot(\fet \tau_2\times\fet
\tau_3)\vec{\sigma}_1\cdot\vec{q}_1\vec{\sigma}_3\cdot\vec{q}_3\vec{\sigma}_2\cdot(\vec{q}_1\times\vec{q}_3)
\quad$ & \quad $\tilde  {\cal G}_{20}  =      \fet \tau_1\cdot(\fet \tau_2\times\fet
\tau_3)\, \vec{\sigma}_1\cdot\hat{r}_{23}\,
\vec{\sigma}_3\cdot\hat{r}_{12}\,
\vec{\sigma}_2\cdot(\hat{r}_{12}\times\hat{r}_{23})$ \,\,\\
\hline
\end{tabular}
\caption{The set of 20 generating operators ${\cal G}_i$ which
  generate $80$ independent operators $O_i$ of a local three-nucleon force.
\label{generalstr}}
\end{table}
For the sake of completeness, we also provide relations between the
old and new structure functions ${\cal F}_i$:
In order to distinguish the new basis from old one, we from now on label a
set of the previous $22$ operators and structure functions by
``old''. We use the relation 
\beq
\sum_{i=1}^{22}{\cal G}_i^{{\rm old}} {\cal F}_i^{{\rm old}}(q_1,q_2,q_3)+
5\,{\rm permutations} = \sum_{i=1}^{20}{\cal G}_i {\cal F}_i(q_1,q_2,q_3)+
5\,{\rm permutations}
\eeq
to express ${\cal F}_i$ in terms of ${\cal F}_i^{{\rm old}}$ via
\beqa
\label{relationsF}
{\cal F}_i(q_1,q_2,q_3)&=&{\cal F}_i^{{\rm old}}(q_1,q_2,q_3) \quad
{\rm for}\quad i=1,\ldots,5,\, 7\ldots 17, \nn
{\cal F}_6(q_1,q_2,q_3)&=&{\cal F}_6^{{\rm old}} (q_1,q_2,q_3)+\Bigg(\frac{1}{24}
q_{1}^2 \left(q_{1}^2-q_{2}^2-3 q_{3}^2\right) 
{\cal F}_{20}^{{\rm old}}(q_{1},q_{2},q_{3})\nn
&+&\frac{1}{24} q_{2}^2 \left(-3 
q_{1}^2+q_{2}^2-3 q_{3}^2\right) {\cal F}_{21}^{{\rm
  old}}(q_{1},q_{2},q_{3})+5\,{\rm permutations}\Bigg),\nn
{\cal F}_{18}(q_1,q_2,q_3)&=&{\cal F}_{18}^{{\rm old}}(q_1,q_2,q_3)+\frac{1}{8} \left(-q_{1}^2+q_{2}^2-q_{3}^2\right) {\cal F}_{20}^{{\rm 
old}}(q_{1},q_{2},q_{3})+\frac{1}{8} \left(q_{1}^2+q_{2}^2-q_{3}^2
\right) {\cal F}_{20}^{{\rm old}}(q_{1},q_{3},q_{2})\nn
&+&\frac{1}{8} 
\left(-q_{1}^2+q_{2}^2+q_{3}^2\right) {\cal F}_{20}^{{\rm 
old}}(q_{3},q_{1},q_{2})+\frac{1}{8} \left(-q_{1}^2+q_{2}^2-q_{3}^2
\right) {\cal F}_{20}^{{\rm old}}(q_{3},q_{2},q_{1})\nn
&+&\frac{1}{8} 
\left(-q_{1}^2+q_{2}^2-q_{3}^2\right) {\cal F}_{21}^{{\rm 
old}}(q_{1},q_{3},q_{2})+\frac{1}{8} \left(-q_{1}^2+q_{2}^2-q_{3}^2
\right) {\cal F}_{21}^{{\rm old}}(q_{2},q_{1},q_{3})\nn
&+&\frac{1}{8} 
\left(-q_{1}^2+q_{2}^2-q_{3}^2\right) {\cal F}_{21}^{{\rm 
old}}(q_{2},q_{3},q_{1})+\frac{1}{8} \left(-q_{1}^2+q_{2}^2-q_{3}^2
\right) {\cal F}_{21}^{{\rm old}}(q_{3},q_{1},q_{2})\nn
&+&\frac{1}{4} q_{2}^2 
{\cal F}_{21}^{{\rm old}}(q_{1},q_{2},q_{3})+\frac{1}{4} q_{2}^2 {\cal F}_{21}^{{
\rm old}}(q_{3},q_{2},q_{1}),\nn
{\cal F}_{19}(q_1,q_2,q_3)&=&{\cal F}_{19}^{{\rm old}}(q_1,q_2,q_3)+\frac{1}{4} \left(-q_{1}^2+q_{2}^2+3 q_{3}^2\right) {\cal F}_{20}^{{\rm 
old}}(q_{1},q_{2},q_{3})+\frac{1}{4} \left(-q_{1}^2+3 q_{2}^2+q_{3}^2
\right) {\cal F}_{20}^{{\rm old}}(q_{1},q_{3},q_{2})\nn
&+&\frac{1}{4} 
\left(q_{1}^2-q_{2}^2-q_{3}^2\right) {\cal F}_{20}^{{\rm 
old}}(q_{3},q_{1},q_{2})+\frac{1}{4} \left(q_{1}^2-q_{2}^2+q_{3}^2
\right) {\cal F}_{20}^{{\rm old}}(q_{3},q_{2},q_{1})\nn
&+&\frac{1}{4} 
\left(q_{1}^2-q_{2}^2+q_{3}^2\right) {\cal F}_{21}^{{\rm 
old}}(q_{1},q_{3},q_{2})+\frac{1}{4} \left(-q_{1}^2+q_{2}^2+3 q_{3}^2
\right) {\cal F}_{21}^{{\rm old}}(q_{2},q_{1},q_{3})\nn
&+&\frac{1}{4} 
\left(q_{1}^2-q_{2}^2+q_{3}^2\right) {\cal F}_{21}^{{\rm 
old}}(q_{2},q_{3},q_{1})+\frac{1}{4} \left(-q_{1}^2+q_{2}^2+3 q_{3}^2
\right) {\cal F}_{21}^{{\rm old}}(q_{3},q_{1},q_{2})\nn
&+&\frac{1}{2} q_{2}^2 
{\cal F}_{21}^{{\rm old}}(q_{1},q_{2},q_{3})+\frac{1}{2} q_{2}^2 {\cal F}_{21}^{{
\rm old}}(q_{3},q_{2},q_{1}),\nn
{\cal F}_{20}(q_1,q_2,q_3)&=&{\cal F}_{22}^{{\rm old}}(q_1,q_2,q_3)-\frac{1}{2} {\cal F}_{20}^{{\rm old}}(q_{1},q_{2},q_{3})-\frac{1}{2} 
{\cal F}_{20}^{{\rm old}}(q_{1},q_{3},q_{2})-\frac{1}{2} {\cal F}_{20}^{{\rm 
old}}(q_{3},q_{1},q_{2})-\frac{1}{2} {\cal F}_{20}^{{\rm 
old}}(q_{3},q_{2},q_{1})\nn
&-&\frac{1}{2} {\cal F}_{21}^{{\rm 
old}}(q_{1},q_{3},q_{2})-\frac{1}{2} {\cal F}_{21}^{{\rm 
old}}(q_{2},q_{1},q_{3})-\frac{1}{2} {\cal F}_{21}^{{\rm 
old}}(q_{2},q_{3},q_{1})-\frac{1}{2} {\cal F}_{21}^{{\rm 
old}}(q_{3},q_{1},q_{2}).
\eeqa
Analogously, in coordinate space we have
\beq
\sum_{i=1}^{22}\tilde{\cal G}_i^{{\rm old}} {\cal F}_i^{{\rm old}}(r_{12},r_{23},r_{31})+
5\,{\rm permutations} = \sum_{i=1}^{20}\tilde{\cal G}_i  {\cal F}_i(r_{12},r_{23},r_{31})+
5\,{\rm permutations}\,,
\eeq
so that $  {\cal F}_i$ can be expressed in terms of $  {\cal F}_i^{{\rm old}}$ via
\beqa
  {\cal F}_i(r_{12},r_{23},r_{31})&=&  {\cal F}_i^{{\rm
      old}}(r_{12},r_{23},r_{31}) \quad {\rm for}\quad i=1\ldots 5, \;
  7
  \ldots 17, \nn
  {\cal F}_6(r_{12},r_{23},r_{31})&=&  {\cal F}_{6}^{{\rm
    old}}(r_{12},r_{23},r_{31}) + \Bigg(\frac{1}{24} r_{23}^2 \left(3
  r_{12}^2+r_{31}^2-r_{23}^2\right) 
  {\cal F}_{20}^{{\rm old}}(r_{12},r_{23},r_{31})\nn
&+&\frac{1}{24} r_{31}^2 
\left(3 r_{12}^2-r_{31}^2+3 r_{23}^2\right)   {\cal F}_{21}^{{\rm 
old}}(r_{12},r_{23},r_{31})+5\,{\rm permutations}\Bigg),\nn
  {\cal F}_{18}(r_{12},r_{23},r_{31})&=&  {\cal F}_{18}^{{\rm
    old}}(r_{12},r_{23},r_{31}) + \frac{1}{8} \left(-r_{12}^2-r_{23}^2+r_{31}^2\right)   {\cal F}_{20}^{{
\rm old}}(r_{12},r_{23},r_{31})\nn
&+&\frac{1}{8} 
\left(-r_{12}^2-r_{23}^2+r_{31}^2\right)   {\cal F}_{20}^{{\rm 
old}}(r_{23},r_{12},r_{31})+\frac{1}{8} 
\left(r_{12}^2-r_{23}^2+r_{31}^2\right)   {\cal F}_{20}^{{\rm 
old}}(r_{31},r_{12},r_{23})\nn
&+&\frac{1}{8} 
\left(-r_{12}^2+r_{23}^2+r_{31}^2\right)   {\cal F}_{20}^{{\rm 
old}}(r_{31},r_{23},r_{12})+\frac{1}{8} 
\left(-r_{12}^2-r_{23}^2+r_{31}^2\right)   {\cal F}_{21}^{{\rm 
old}}(r_{12},r_{31},r_{23})\nn
&+&\frac{1}{8} 
\left(-r_{12}^2-r_{23}^2+r_{31}^2\right)   {\cal F}_{21}^{{\rm 
old}}(r_{23},r_{31},r_{12})+\frac{1}{8} 
\left(-r_{12}^2-r_{23}^2+r_{31}^2\right)   {\cal F}_{21}^{{\rm 
old}}(r_{31},r_{12},r_{23})\nn
&+&\frac{1}{8} 
\left(-r_{12}^2-r_{23}^2+r_{31}^2\right)   {\cal F}_{21}^{{\rm 
old}}(r_{31},r_{23},r_{12})+\frac{1}{4} r_{31}^2   {\cal F}_{21}^{{\rm 
old}}(r_{12},r_{23},r_{31})+\frac{1}{4} r_{31}^2   {\cal F}_{21}^{{\rm 
old}}(r_{23},r_{12},r_{31}),\nn
  {\cal F}_{19}(r_{12},r_{23},r_{31})&=&  {\cal F}_{19}^{{\rm
    old}}(r_{12},r_{23},r_{31}) +\frac{1}{4} \left(-3 r_{12}^2+r_{23}^2-r_{31}^2\right)   
{\cal F}_{20}^{{\rm old}}(r_{12},r_{23},r_{31})\nn
&+&\frac{1}{4} 
\left(-r_{12}^2-r_{23}^2+r_{31}^2\right)   {\cal F}_{20}^{{\rm 
old}}(r_{23},r_{12},r_{31})+\frac{1}{4} 
\left(r_{12}^2-r_{23}^2+r_{31}^2\right)   {\cal F}_{20}^{{\rm 
old}}(r_{31},r_{12},r_{23})\nn
&+&\frac{1}{4} \left(-r_{12}^2+r_{23}^2-3 
r_{31}^2\right)   {\cal F}_{20}^{{\rm 
old}}(r_{31},r_{23},r_{12})+\frac{1}{4} \left(-3 
r_{12}^2+r_{23}^2-r_{31}^2\right)   {\cal F}_{21}^{{\rm 
old}}(r_{12},r_{31},r_{23})\nn
&+&\frac{1}{4} 
\left(-r_{12}^2-r_{23}^2+r_{31}^2\right)   {\cal F}_{21}^{{\rm 
old}}(r_{23},r_{31},r_{12})+\frac{1}{4} \left(-3 
r_{12}^2+r_{23}^2-r_{31}^2\right)   {\cal F}_{21}^{{\rm 
old}}(r_{31},r_{12},r_{23})\nn
&+&\frac{1}{4} 
\left(-r_{12}^2-r_{23}^2+r_{31}^2\right)   {\cal F}_{21}^{{\rm 
old}}(r_{31},r_{23},r_{12})-\frac{1}{2} r_{31}^2   {\cal F}_{21}^{{\rm 
old}}(r_{12},r_{23},r_{31})-\frac{1}{2} r_{31}^2   {\cal F}_{21}^{{\rm 
old}}(r_{23},r_{12},r_{31}),\nn
  {\cal F}_{20}(r_{12},r_{23},r_{31})&=&  {\cal F}_{22}^{{\rm
    old}}(r_{12},r_{23},r_{31}) -\frac{1}{2}   {\cal F}_{20}^{{\rm 
old}}(r_{12},r_{23},r_{31})-\frac{1}{2}   {\cal F}_{20}^{{\rm 
old}}(r_{23},r_{12},r_{31})-\frac{1}{2}   {\cal F}_{20}^{{\rm 
old}}(r_{31},r_{12},r_{23})\nn
&-&\frac{1}{2}   {\cal F}_{20}^{{\rm 
old}}(r_{31},r_{23},r_{12})-\frac{1}{2}   {\cal F}_{21}^{{\rm 
old}}(r_{12},r_{31},r_{23})-\frac{1}{2}   {\cal F}_{21}^{{\rm 
old}}(r_{23},r_{31},r_{12})-\frac{1}{2}   {\cal F}_{21}^{{\rm 
old}}(r_{31},r_{12},r_{23})\nn
&-&\frac{1}{2}   {\cal F}_{21}^{{\rm 
old}}(r_{31},r_{23},r_{12}).
\eeqa

\section{Chiral expansion of the three-nucleon force in coordinate space}
\def\theequation{\arabic{section}.\arabic{equation}}
\label{sec:3nfcoordspace}

We are now in the position to discuss the contributions of the long-
and intermediate-range 3NF topologies to the structure functions
${\cal F}_{i} (r_{31},r_{23},r_{12})$. 

We begin with the longest-range $2\pi$-exchange 3NF whose explicit
expressions at N$^2$LO, N$^3$LO and N$^4$LO are given in Ref.~\cite{Krebs:2012yv}
both in momentum and coordinate spaces. Following the lines of
Ref.~\cite{Krebs:2013kha},  
we restrict ourselves in this qualitative discussion to the equilateral
triangle configuration with $r_{12} = r_{23} = r_{31} \equiv r$ which
allows us to visualize the structure functions in a simple way. Notice
that while this is sufficient for  a qualitative estimation of the
size of various contributions, the final
conclusions about the importance of the individual structures in the
3NF for nuclear observables can only be drawn upon solving the
quantum-mechanical $A$-body problem. Work along these lines is in
progress, see \cite{Witala:2013kya,Golak:2014ksa}  for some preliminary results. 

\begin{figure}[tb]
\vskip 1 true cm
\includegraphics[width=\textwidth,keepaspectratio,angle=0,clip]{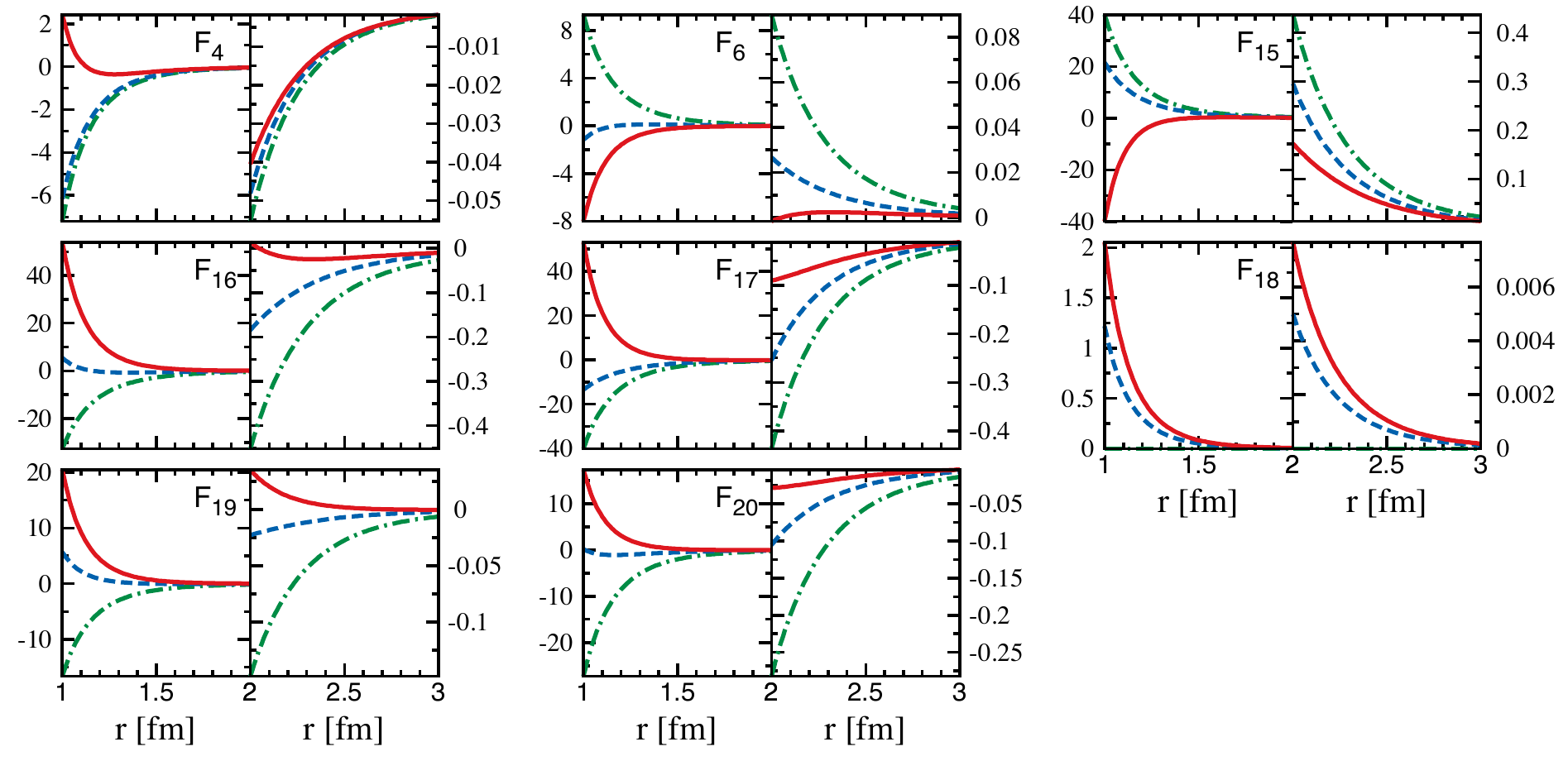}
    \caption{Chiral expansion of the profile functions ${\cal F}_i
      (r)$ in MeV generated by the two-pion exchange 3NF topology up to
      N$^4$LO (in the equilateral triangle configuration). Dashed-dotted, dashed and solid lines correspond to
      ${\cal F}_i^{(3)}$,   ${\cal F}_i^{(3)} + {\cal F}_i^{(4)}$ and
      ${\cal F}_i^{(3)} + {\cal F}_i^{(4)} + {\cal F}_i^{(5)}$,
      respectively. 
\label{fig:TPEr} 
 }
\end{figure}

In Fig.~\ref{fig:TPEr} we show the chiral expansion of the structure functions ${\cal F}_i
      (r)$ generated by the $2\pi$-exchange 3NF topology up to
      N$^4$LO. 
Here and in what follows, we use the values for the various LECs from
Ref.~\cite{Krebs:2012yv}  corresponding to  the order-$Q^4$ fit to
pion-nucleon phase shifts from the Karsruhe-Helsinki (KH) partial-wave
analysis \cite{Koch:1985bn}. Specifically, we use $M_\pi = 138$ MeV,
$F_\pi=92.4$ MeV, $g_A = 1.285$\footnote{This value takes into account
   the
  Goldberger-Treiman discrepancy.} for the pion mass, pion decay constant
and the nucleon axial vector coupling while the values of the
other relevant  LECs  are: $c_1 = -0.75$ GeV$^{-1}$,   $c_2 = 3.49$
GeV$^{-1}$, $c_3 = -4.77$ GeV$^{-1}$, $c_4 = 3.34$ GeV$^{-1}$,
$\bar e_{14} = -1.52$ GeV$^{-3}$ and $\bar e_{17} = -0.37$ GeV$^{-3}$.
Notice that while the function ${\cal F}_i (r)$ are shown in the range of $1 \ldots
3$ fm, the chiral expansion for the potentials is expected to converge
only at sufficiently large distances, see Ref.~\cite{Baru:2012iv} for a related
discussion.  
The most recent versions of the chiral nucleon-nucleon potentials
employ local regularization of the pion-exchange
contributions  in coordinate space with the cutoff $R_0 \sim 1$ fm
\cite{Gezerlis:2013ipa,Gezerlis:2014zia}. Such a regulator would clearly strongly affect the behavior of the
functions ${\cal F}_i  (r)$ at short distances but would  have little impact
at relative distances  $r > 2$ fm.  

The  $2\pi$-exchange topology gives rise to $8$ out
of $20$ operators.  Surprisingly, one observes that the chiral expansion only
appears to converge at this order for rather large distances beyond $r
\sim 2.5$ fm. At such distances, the N$^4$LO corrections are indeed
considerably smaller than the N$^3$LO ones. In this context, it is
important to keep in mind that contrary to the N$^3$LO corrections,
the N$^4$LO ones  
involve terms proportional to the LECs $c_{2,3,4}$ which receive
contributions from the $\Delta$ isobar and appear to be numerically
large. Thus, one may indeed expect the N$^4$LO contributions to be larger
than what is suggested by naive dimensional analysis which, at least to
some extent, may explain the observed convergence pattern. 
The convergence of the chiral expansion for the
$2\pi$-exchange 3NF was also addressed in Ref.~\cite{Krebs:2012yv} based on the
momentum-space expressions for the function ${\cal A} (q_2)$ and
${\cal B} (q_2)$, which parametrize the pion-nucleon amplitude in the
kinematics relevant to the 3NF. One observes from Fig. 5 of that work that the
N$^4$LO contributions to both of these functions are significantly smaller than
the N$^3$LO ones in the range of momentum transfers of $q_2 <  300$
MeV. Confronting these findings with results in coordinate space
suggests that higher-momentum components do significantly affect the
potential at relative distances of the order of $r \sim 2$ fm, see 
Ref.~\cite{Epelbaum:2003gr} for a related discussion.  

We also observe an interesting feature that the N$^3$LO and N$^4$LO
corrections contribute in the same direction and lead to a strong
reduction in magnitude of the strength of the potentials at distances of the order of 
$r \sim 2$ fm. For example, the
strongest potentials ${\cal F}_{15} (r)$, ${\cal F}_{16} (r)$ and ${\cal
  F}_{17} (r)$ have, at the relative distance $r = 2$ fm, the strength
of $440$ keV, $-450$ keV and $-440$ keV at
N$^2$LO while  $170$ keV, $14$ keV and $-90$ keV at N$^4$LO. This feature,
that the N$^2$LO results based on the $c_i$'s taken from the
order-$Q^4$ fit to pion-nucleon phase shifts tend to strongly
overshoot the $2\pi$-exchange  3NF contribution, is consistent with the
observations of Ref.~\cite{Krebs:2012yv} in momentum space. 

\begin{figure}[tb]
\vskip 1 true cm
\includegraphics[width=\textwidth,keepaspectratio,angle=0,clip]{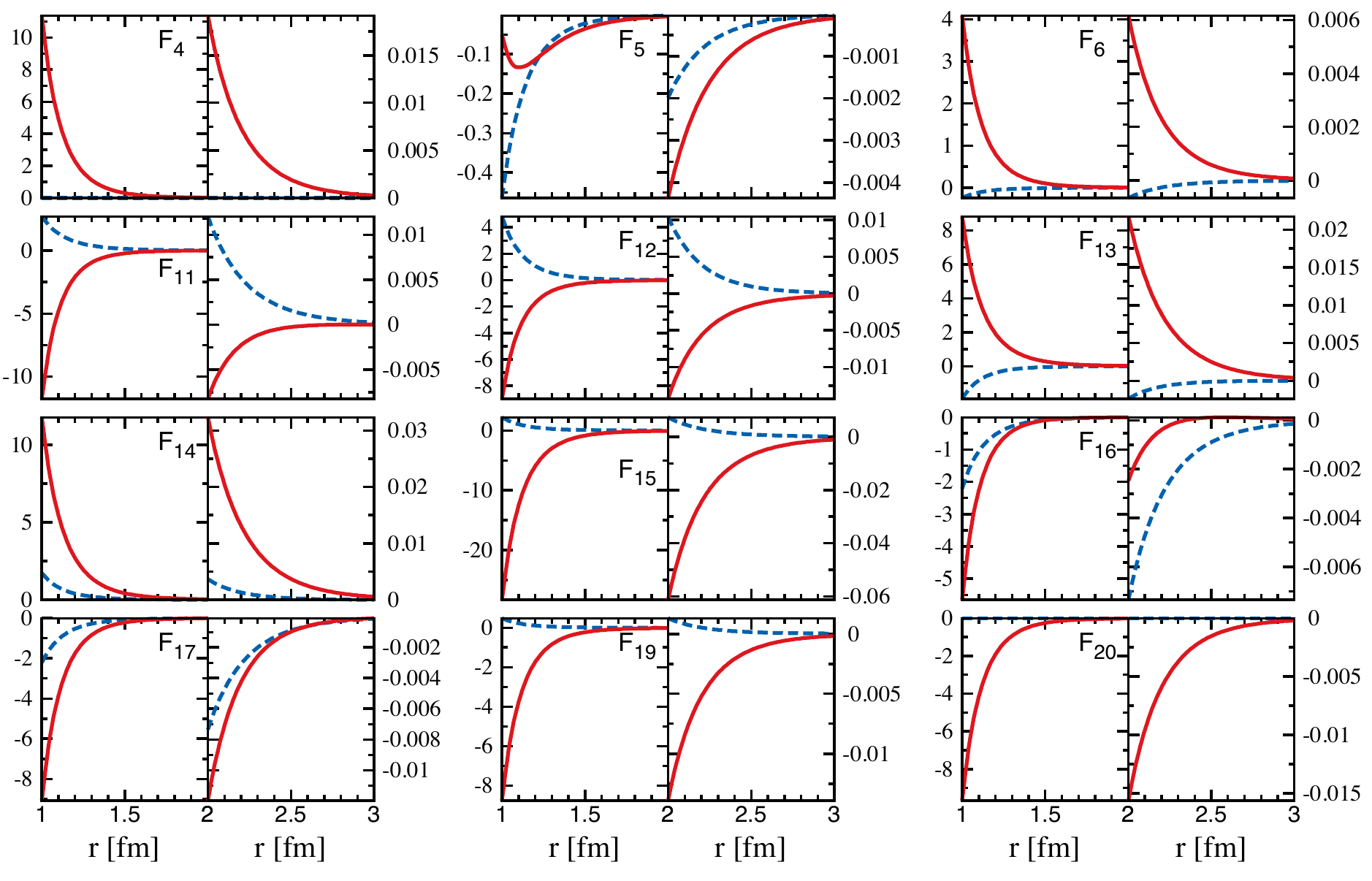}
    \caption{Chiral expansion of the profile functions ${\cal F}_i
      (r)$ in MeV generated by the two-pion-one-pion exchange  3NF topology up to
      N$^4$LO (in the equilateral triangle configuration). Dashed and solid lines correspond to
      ${\cal F}_i^{(4)}$ and  ${\cal F}_i^{(4)} + {\cal F}_i^{(5)}$,
      respectively. 
\label{fig:TPEOPEr} 
 }
\end{figure}

\begin{figure}[tb]
\vskip 1 true cm
\includegraphics[width=\textwidth,keepaspectratio,angle=0,clip]{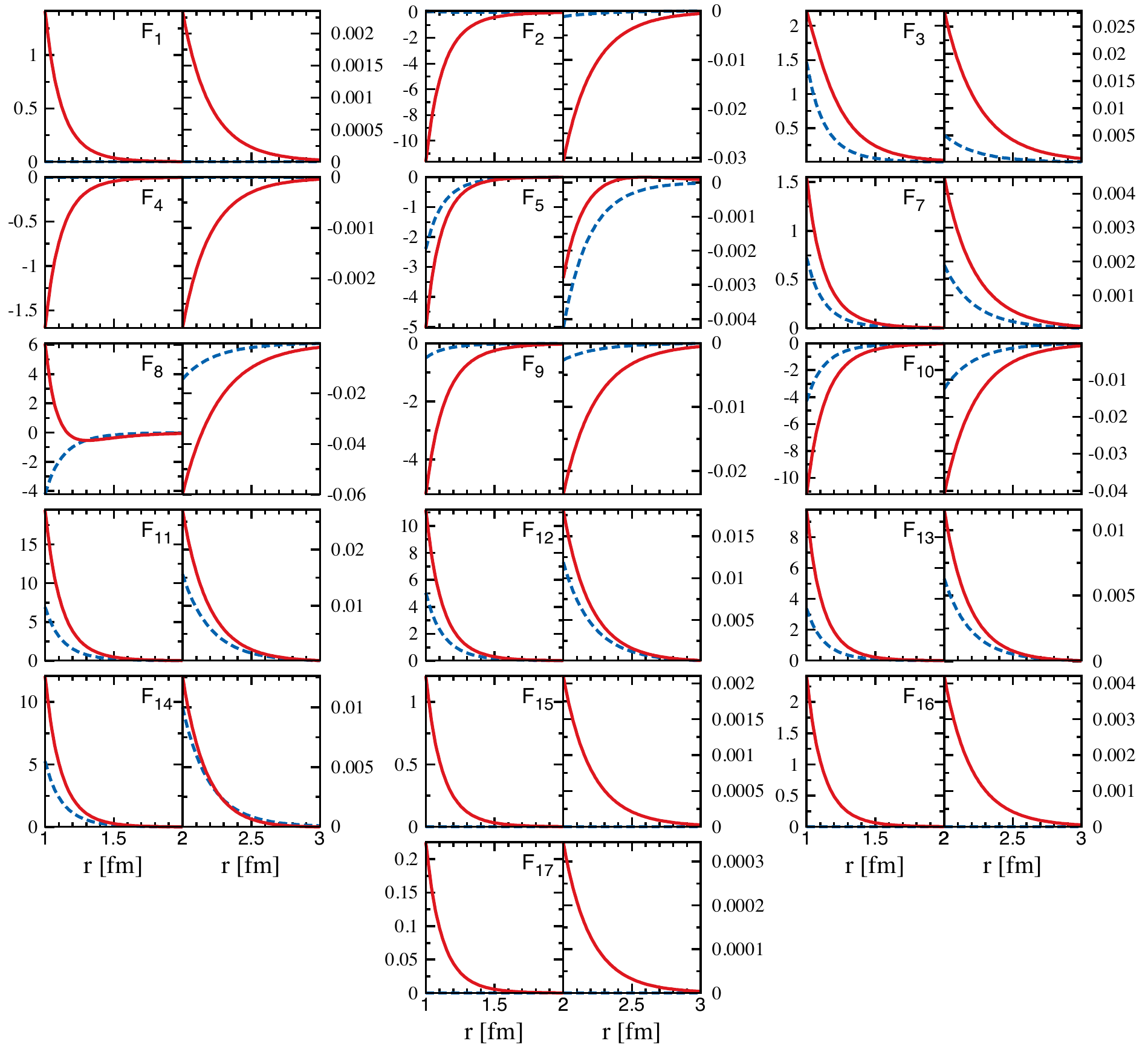}
    \caption{Chiral expansion of the profile functions ${\cal F}_i
      (r)$ in MeV generated by the ring  3NF topology up to
      N$^4$LO (in the equilateral triangle configuration). Dashed and solid lines correspond to
      ${\cal F}_i^{(4)}$ and  ${\cal F}_i^{(4)} + {\cal F}_i^{(5)}$,
      respectively. 
\label{fig:ringr} 
 }
\end{figure}

The results for the $2\pi$-$1\pi$ exchange and ring topologies are
depicted in Figs.~\ref{fig:TPEOPEr} and \ref{fig:ringr}, respectively.     
Given that these are genuine one-loop topologies, the chiral expansion
for these contributions starts at N$^3$LO. Notice further that only the results for 
${\cal F}_{6,19,20}$ are affected by using the new operator
basis, see Eq.~(\ref{relationsF}) for explicit expressions. Thus, all conclusions of Ref.~\cite{Krebs:2013kha} remain
unaffected. In particular, one observes that the N$^4$LO terms are in most cases larger
in magnitude than the (nominally) leading contributions at 
N$^3$LO. This pattern is in line with the assumption that these
contributions are, to a large extent, driven by 
intermediate $\Delta$ excitations. In the $\Delta$-less
formulation of chiral EFT, these effects for the considered 3NF topologies start to appear at N$^4$LO.  

\begin{figure}[tb]
\vskip 1 true cm
\includegraphics[width=\textwidth,keepaspectratio,angle=0,clip]{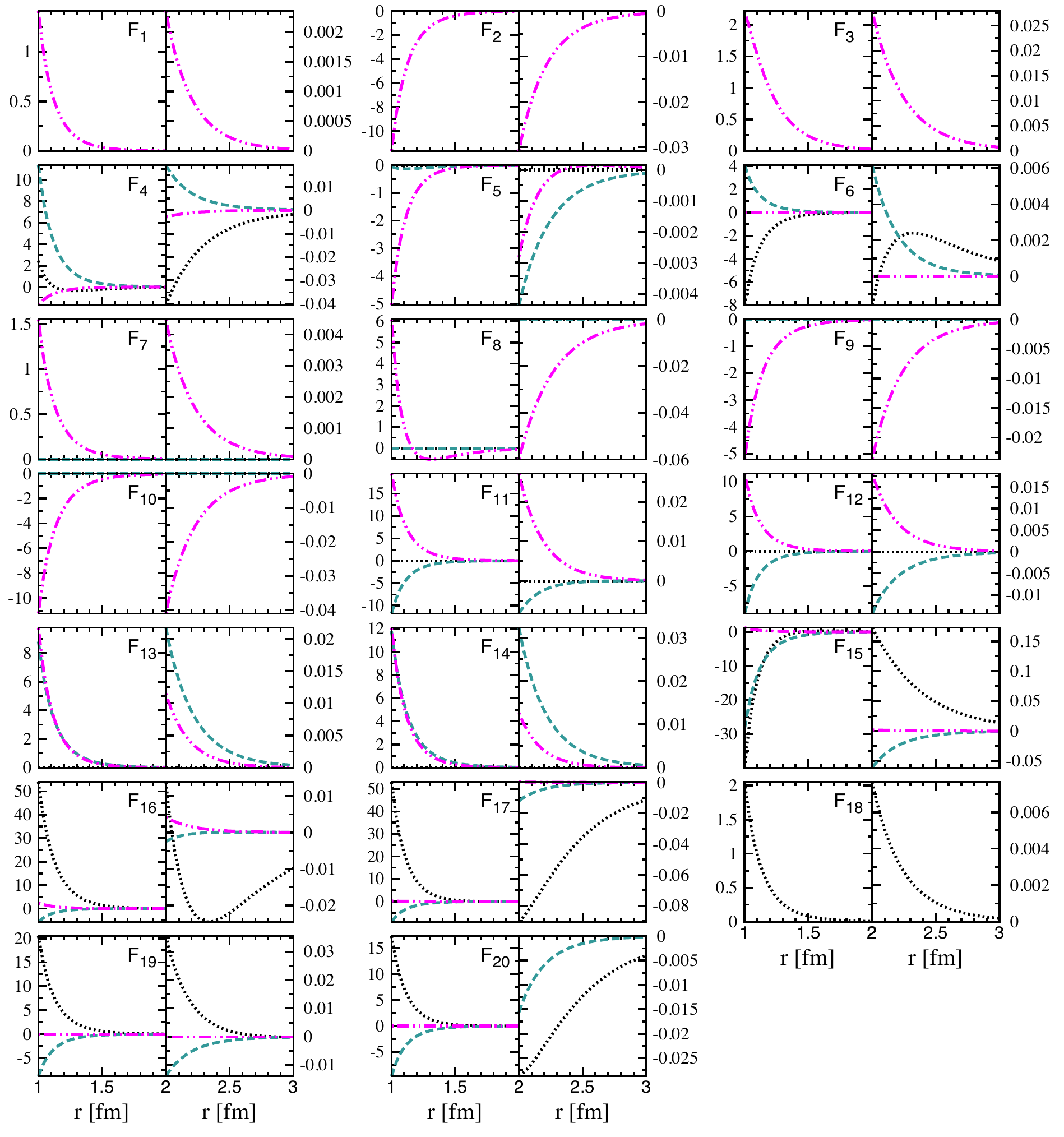}
    \caption{Individual contributions of the two-pion exchange (dotted
      lines), two-pion-one-pion exchange (long-dashed lines) and ring
      (dashed-double-dotted lines) topologies to the profile functions  ${\cal F}_i
      (r)$ in MeV at N$^4$LO in the equilateral triangle configuration. 
\label{fig:topologiesr} 
 }
\end{figure}

It is instructive to compare the potentials at large distances
emerging from the individual topologies with each other. This is
visualized in Fig.~\ref{fig:topologiesr} where only N$^4$LO
results for the  functions ${\cal  F}_i$ are shown. 
One clearly observes the longest-range nature of the $2\pi$-exchange
3NF which, in all cases where it doesn't vanish, dominates the potential
at distances larger than $r = 2$ fm. In particular, the strongest $2\pi$-exchange
potentials  ${\cal F}_{15} (2\mbox{ fm}) \simeq 170$ keV, ${\cal F}_{17}
(2\mbox{ fm}) \simeq -90$ keV are considerably larger in magnitude than the
strongest $2\pi$-$1\pi$   ${\cal F}_{14} (2\mbox{ fm}) \simeq 29$ keV, ${\cal F}_{15}
(2\mbox{ fm}) \simeq -69$ keV and ring potentials ${\cal F}_{8} (2\mbox{ fm}) \simeq -60$ keV, ${\cal F}_{10}
(2\mbox{ fm}) ~\simeq -41$ keV, respectively. This dominance becomes
more pronounced at larger distances while at shorter ones 
all three topologies generate contributions of a comparable size.  We
emphasize once again that more quantitative conclusions about 
importance of individual 3NF contributions can only be
drawn upon performing explicit calculations of few-nucleon observables. 

\begin{figure}[tb]
\vskip 1 true cm
\includegraphics[width=\textwidth,keepaspectratio,angle=0,clip]{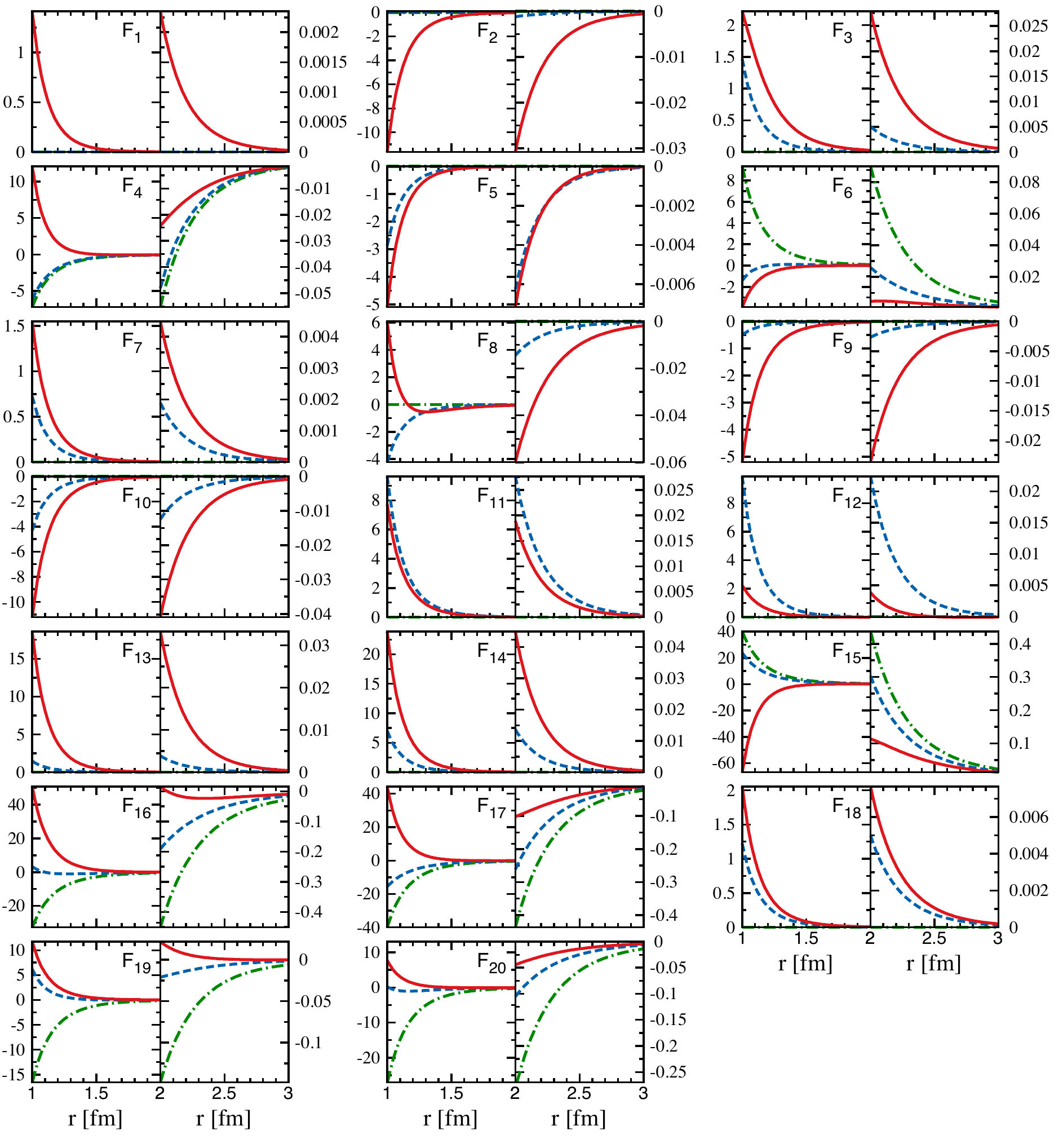}
    \caption{Chiral expansion of the profile functions ${\cal F}_i
      (r)$ in MeV emerging from all long-range 3NF topologies up to
      N$^4$LO (in the equilateral triangle configuration). Dashed-dotted, dashed and solid lines correspond to
      ${\cal F}_i^{(3)}$,   ${\cal F}_i^{(3)} + {\cal F}_i^{(4)}$ and
      ${\cal F}_i^{(3)} + {\cal F}_i^{(4)} + {\cal F}_i^{(5)}$,
      respectively. 
\label{fig:completer} 
 }
\end{figure}

Finally, Fig.~\ref{fig:completer} shows the resulting chiral expansion of the
structure functions ${\cal F}_i$ when all three types of contributions
are added together.  
These plots clearly reflect the behavior observed for individual
topologies as discussed above. 
The
strongest potentials at $r = 2$ fm are ${\cal F}_{15}  \simeq 100$ keV
and ${\cal F}_{17} \simeq -90$ keV, while at $r \sim M_\pi^{-1} \sim 1.4$
fm one has ${\cal F}_{16}  \simeq 2.9$ MeV and ${\cal F}_{17} \simeq 1.4$ MeV.   

As already pointed out in Ref.~\cite{Krebs:2013kha}, given the large
corrections at the subleading one-loop level, i.e. N$^4$LO, which are
driven by single-delta excitations, it is important to  study
contributions emerging from intermediate double- and triple-$\Delta$
excitations.  In the standard $\Delta$-less formulation of chiral EFT,
such contributions first appear at N$^5$LO and N$^6$LO, respectively, where one would
also need to evaluate all possible two-loop diagrams. This is clearly
a rather challenging task. A more
promising and feasible approach would be to employ the formulation
of EFT with explicit $\Delta$ degrees of freedom. Such a framework was
shown in the past to be quite efficient in resumming the large
contributions to the nuclear force associated with intermediate $\Delta$
excitations \cite{Ordonez:1993tn,Kaiser:1998wa,Krebs:2007rh,Epelbaum:2007sq,Epelbaum:2008td}. In this formulation, effects of single-, double- and
triple-$\Delta$ excitations are accounted for at the leading one-loop
level, i.e. N$^3$LO. Work along these lines is in progress.

\section{Large-$N_c$ insights}
\def\theequation{\arabic{section}.\arabic{equation}}
\label{sec:largeNC}

It is interesting to analyze our findings  for the profile functions in the light
of the $1/N_c$-expansion of QCD. The final results summarized in
Fig.~\ref{fig:completer} show clearly that not all profile functions are of the
same size. In particular, at $r \sim 1$ fm, the absolute values of the
functions $|{\cal F}_{15,16,17}| \gtrsim 40 \; \rm MeV$ appear to be
much larger than the other ${\cal F}_i$'s for no apparent
reason. As we will argue below, this pattern is in line with the
large-$N_c$ picture of the 3NF. 

The large-$N_c$ expansion of QCD  proved to be a useful
approach for understanding various qualitative aspects of mesons and
baryons \cite{'tHooft:1973jz,Witten:1979kh}, see Ref.~\cite{Jenkins:1998wy} for a review article. In
particular, it 
was applied in Refs.~\cite{Kaplan:1995yg,Kaplan:1996rk} to explain 
the pattern in the relative strengths of various
spin-flavor components of the nucleon-nucleon force as observed in 
phenomenological models. Recently, these studies were extended to 
the 3NF \cite{Phillips:2013rsa} by classifying the operators appearing in the 
3NF according to their large-$N_c$ scaling.  It is thus interesting to
confront these insights with the chiral EFT calculations presented in this
work. We recall that according to the analysis of
Ref.~\cite{Phillips:2013rsa},  the
operators ${\cal G}_{1,4,6,15,16,17,18,19,20}$ appear at leading order
$\mathcal{O}(N_c)$ while all other  structures in
Table~\ref{generalstr} appear at subleading
order $\mathcal{O}(1/N_c)$.  Thus, the observed numerical dominance of $|{\cal F}_{15,16,17}|$ is consistent with
the corresponding operators contributing at leading order in the large-$N_c$ scaling. While some of the other profile functions
for the order-$\mathcal{O}(N_c)$ structures come out smaller,  this does not imply a violation
of the $N_c$-scaling. 

\begin{figure}[tb]
\vskip 1 true cm
\includegraphics[width=8cm,keepaspectratio,angle=0,clip]{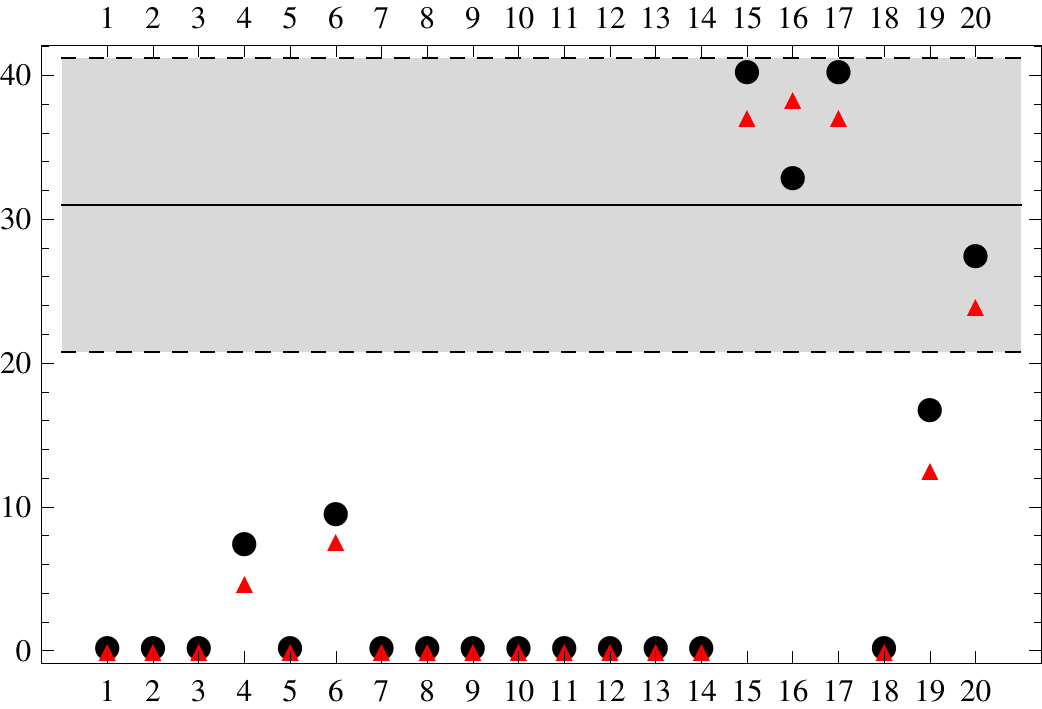}
\qquad
\qquad
\includegraphics[width=8cm,keepaspectratio,angle=0,clip]{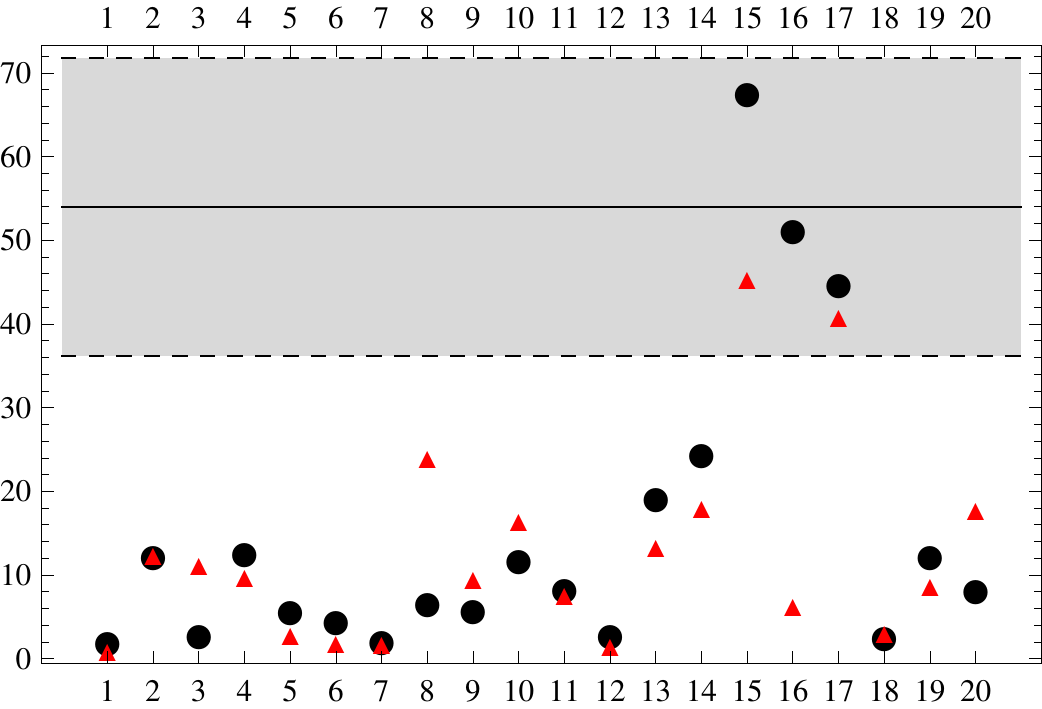}
	\caption{Absolute values of the profile functions ${\cal
            F}_i$,  with $i=1, \ldots 20$ shown on the $x$ axis,  
          in the equilateral triangle configuration at the relative
          distance of  $r=1$ fm
          (black dots)  and $r=2$ fm (red triangles) at N$^2$LO  in
          the left
          panel and at N$^4$LO in the right panel. The values at $r=1$ fm
          are shown in units of MeV while the ones at $r=2$ fm are in
          units of $1/85$ MeV (left panel) and  $1/400$ MeV (right
          panel). The gray band corresponds to a 30 \% uncertainty due
to subleading $1/N_c$-corrections. } 
\label{Fsize}
\end{figure}

In order to get further insights into the hierarchy of various profile
functions, we plot in Fig.~\ref{Fsize} the absolute values of the
corresponding potentials in the equilateral triangle configuration at distances of $r=1$ fm and $r=2$ fm at
N$^2$LO (left panel) and N$^4$LO (right panel). Here, the horizontal
black line is the average of $|{\cal F}_{15,16,17,19,20}|$
(left panel) and  $|{\cal F}_{15,16,17}|$ (right panel)  at $r=1$ fm  and serves as
an estimation of the natural size of $|{\cal F}^{(3)}_i |$ and $|{\cal
  F}^{(3)}_i + {\cal
  F}^{(4)}_i  + {\cal
  F}^{(5)}_i |$ at that distance.
At N$^2$LO, the
observed hierarchy of the various flavor-spin-space structures in the
3NF is in a good agreement with the  expected pattern based on the large-$N_c$
analysis. In particular,  except for ${\cal F}_1$ and ${\cal F}_{18}$,  all
profile functions corresponding to the leading in the $1/N_c$-counting
structures receive sizable contributions at N$^2$LO. This should not
come as a surprise: indeed, it was shown in
Ref.~\cite{Phillips:2013rsa} that the dominant, i.e.~order $\sim N_c$, 3NF
contains all operators present in the Fujita-Miyazawa 3NF model
\cite{Fujita:1957zz}, which has the same structure as the leading chiral
$2\pi$-exchange 3NF, see Eqs.~(3.3) and (3.5) of
Ref.~\cite{Krebs:2012yv}. In fact, given that $g_A \sim \mathcal{O}
(N_c)$ and $F_\pi \sim  \mathcal{O}
(N_c^{1/2})$, one can immediately read off from these
expressions that the  N$^2$LO $2\pi$-exchange 3NF is of order
$\mathcal{O} (N_c)$ in the regime
of $| \vec p_i \, | \sim \mathcal{O}
(N_c^{0})$. Here we assumed that the LECs $c_i$
scale as $c_i \sim \mathcal{O} (N_c)$ which can be verified
e.g.~within the resonance saturation picture as discussed in
Ref.~\cite{Bernard:1996gq}.\footnote{Notice that for the sake of
  the large-$N_c$ estimations of the LECs $c_{2,3,4}$, it is not
  legitimate to employ the expansion in powers of $M_\pi/(m_\Delta -
  m_N)$ as done in $\Delta$-less formulations of chiral effective field theory.} 
Notice further that the eight coordinate-space profile functions for the
$2\pi$-exchange 3NF discussed above emerge from just 
two flavor-spin-momentum structures ${\cal G}_{17,20}$ upon making 
a Fourier transformation. 

At $\rm N^4LO$, the profile functions show a qualitatively similar
pattern to the one observed at N$^2$LO and discussed above, but the
picture is not that clear anymore. While the strongest potentials are
still the ones corresponding to the operators $\tilde {\cal
  G}_{15,16,17}$ which appear at leading order $\mathcal{O}(N_c)$, the
remaining weaker potentials show no clear pattern with respect to the
large-$N_c$ counting.  It should, however, be emphasized that beyond $\rm
N^2LO$, higher-order diagrams with a larger number of vertices scale
with increasingly higher powers of $N_c$, which naively seems to destroy the $1/N_c$
hierarchy. As it is well known
\cite{Jenkins:1995gc,FloresMendieta:2000mz}, consistency requires
delicate cancellations from $\Delta$ intermediate states, that at the one loop
level and in the one-nucleon sector are currently subject to investigation
\cite{CalleCordon:2012xz}.  Large-$N_c$ consistency was also verified
within the boson-exchange picture of the nucleon-nucleon interaction in
Refs.~\cite{Banerjee:2001js,Belitsky:2002ni,Cohen:2002im} at the 
three-meson exchange level provided the potential is
defined in a specific way. It remains to be seen whether the large-$N_c$ consistency
holds true for nuclear potentials defined with the method of unitary
transformation \cite{Epelbaum:1998ka,Epelbaum:1999dj}. Irrespective of this issue, we further
emphasize that the large-$N_c$ insights into nuclear forces of 
Refs.~\cite{Phillips:2013rsa,Kaplan:1995yg,Kaplan:1996rk,Banerjee:2001js,Belitsky:2002ni,Cohen:2002im}
are achieved assuming the regime in which typical momenta of the
nucleons are $|
\vec p \, | \sim \mathcal{O}(N_c^0)$, and the 
$\Delta$-isobar has to be  treated as an explicit degree of freedom since $m_\Delta - m_N
\sim \mathcal{O} (N_c^{-1})$. These conditions
differ substantially from the ones underlying our chiral EFT
calculations where, in particular, we assign $| \vec p \, | \sim M_\pi \ll m_\Delta - m_N$. 
It is conceivable that the impact of this mismatch increases with
increasing the chiral order so that the comparison between the two
approaches beyond N$^2$LO should be taken with care.

\section{Summary and outlook}
\def\theequation{\arabic{section}.\arabic{equation}}
\label{sec:summary}

The pertinent results of our study can be summarized as follows:
\begin{itemize}
\item
We have clarified the issue with the different number of operators
needed to parametrize the most general isospin-invariant local 3NF
reported in Refs.~\cite{Phillips:2013rsa,Krebs:2013kha}. In particular, we have shown that $2$ out of $22$
operators listed in Ref.~\cite{Krebs:2013kha} are redundant so that the operator basis
involves $20$ flavor-spin-space or, equivalently, flavor-spin-momentum
operators. This agrees with the findings of Ref.~\cite{Phillips:2013rsa}. We also provided explicit
expressions which can be used to rewrite the two redundant structures
in terms of the remaining $20$ operators. 
\item
We re-considered the results for the long- and intermediate-range 3NF
up to N$^4$LO of Ref.~\cite{Krebs:2013kha} using this new operator basis. In
particular, we discussed in detail the convergence of the chiral
expansion for the corresponding profile functions in the equilateral
triangle topology. As expected, we found large N$^4$LO
contributions to the $2\pi$-$1\pi$ exchange and ring
topologies. Moreover, somewhat surprisingly, the  N$^4$LO corrections
to the longest-range $2\pi$ exchange topology are found to be still
sizable even at relatively large distances. Furthermore, we found that
taking into
account N$^3$LO and N$^4$LO corrections to the  $2\pi$ exchange 3NF
amounts to a considerable reduction of  the strength of nearly all profile functions at large
distances and thus makes the 3NF more short-ranged. 
Comparing the potentials generated by the individual
topologies with each other, we observe a clear dominance of the
longest range $2\pi$ exchange at  distances of $r > 2 $ fm, while at
short distances of $r \sim 1$ fm the contributions of $2\pi$-$1\pi$
and ring graphs start becoming
comparable in size. We also see that the $2\pi$-$1\pi$ exchange and
the ring topologies generate sizable intermediate-range potentials 
in those structures where the $2\pi$ exchange does not
contribute. 
\item
We found that the obtained results for the longest- and
intermediate-range topologies agree at the qualitative level with the
results of the large-$N_c$ analysis of Ref.~\cite{Phillips:2013rsa}. We argued
that a more quantitative comparison between the two approaches 
might be difficult due to the different kinematical regimes assumed in 
the two methods. 
\end{itemize}
The present study represents an important intermediate step towards
high-precision analysis of the 3NF in chiral EFT and should be
extended in different ways. First, one needs to work out the
remaining one-pion-exchange-contact and two-pion-exchange-contact contributions to the
3NF. Together with the results reported in
Refs.~\cite{Krebs:2013kha,Krebs:2012yv,Girlanda:2011fh}, 
this will provide a complete  representation of the 3NF
at N$^4$LO. Independently of these studies, one should analyze the 3NF
at N$^3$LO employing the formulation of chiral EFT where the $\Delta$-isobar
is explicitly taken into account
\cite{Ordonez:1993tn,Kaiser:1998wa,Krebs:2007rh,Epelbaum:2007sq}.  A detailed comparison between the two
approaches will shed light on the convergence of the chiral expansion
and allow one to draw conclusions about the size of higher-order terms and
delta-contributions. Finally and most importantly, the resulting novel
terms in the 3NF should be partial wave decomposed \cite{Golak:2009ri} and 
employed in \emph{ab-initio} few- and many-body calculations of
nuclear reactions and light nuclei, see Refs.~\cite{Witala:2013kya,Golak:2014ksa} for first steps
in that direction. Work along these lines is in progress.

\section*{Acknowledgments}

This work is supported by the EU HadronPhysics3 project ``Study of strongly interacting matter'', 
by the European Research Council (ERC-2010-StG 259218 NuclearEFT) and 
by the DFG (TR 16, ``Subnuclear Structure of Matter'').

\end{document}